\documentstyle[12pt]{article}
\newcommand{\pv}{\mbox{$\vec{p}$}}
\newcommand{\qv}{\mbox{$\vec{q}$}}
\newcommand{\kv}{\mbox{$\vec{k}$}}
\newcommand{\lv}{\mbox{$\vec{l}$}}
\newcommand{\xv}{\mbox{$\vec{x}$}}
\newcommand{\yv}{\mbox{$\vec{y}$}}
\newcommand{\Xv}{\mbox{$\vec{X}$}}
\newcommand{\Yv}{\mbox{$\vec{Y}$}}
\newcommand{\kprop}{\mbox{$\vec{k^2}-m^2 +i\epsilon$}}

\newcommand{\pprop}{\mbox{$\vec{p^2}-m^2 +i\epsilon$}}
\newcommand{\qprop}{\mbox{$\vec{q^2}-m^2 +i\epsilon$}}
\newcommand{\kqprop}{\mbox{$q_0^2-k^2-m^2 +i\epsilon$}}
\begin{document}
\baselineskip=22pt plus 0.2pt minus 0.2pt
\lineskip=22pt plus 0.2pt minus 0.2pt
\font\bigbf=cmbx10 scaled\magstep3
\begin{center}
 {\bigbf Path integral quantization of parametrised field theory}\\

\vspace*{0.35in}

\large

Madhavan Varadarajan
\vspace*{0.25in}

\normalsize

{\sl Raman Research Institute,
Bangalore 560 080, India.}
\\
madhavan@rri.res.in\\
\vspace{.5in}

March 2004\\
\vspace{.5in}
ABSTRACT: Free scalar field theory on a flat spacetime can be cast into a generally covariant form known as parametrised field theory in which the action is a functional of the scalar field as well as the embedding variables which describe arbitrary , in general curved, foliations of the flat spacetime. We construct the path integral quantization of parametrised field theory in order to analyse
issues at the interface of quantum field theory and general covariance in a 
path integral context. We show that the  measure in the Lorentzian 
path integral
is non-trivial and is the analog of the Fradkin-Vilkovisky measure for
quantum gravity. 
We construct Euclidean functional integrals in the generally covariant setting of parametrised field theory using key ideas of Schleich and show that  our constructions imply the existence of non-standard `Wick rotations' of the standard
free scalar field 2 point function. We develop a framework to study the problem
of time through computations of scalar field 2 point functions. We illustrate
our ideas through explicit computation for a time independent 
1+1 dimensional foliation.
Although the problem of time seems to be absent in this simple example, the 
general case is still open. 
We discuss our results in the contexts
of the path integral formulation of quantum gravity and 
the canonical quantization
of parametrised field theory.

\end{center}

\pagebreak

\setcounter{page}{1}

\section*{1. Introduction} 

Most treatments of quantum fields on a flat spacetime are based on the 
existence of foliations of the spacetime by flat slices of constant inertial 
time. In generally covariant systems like general relativity, no preferred foliations exist.
Indeed, general covariance requires that all spacelike foliations should be allowed in the description of dynamics. This is only one aspect of the many new conceptual and technical issues which arise in attempts to quantize the gravitational field. In order to isolate and understand  this aspect better, it is useful to study quantum field theory on {\em curved} foliations of flat spacetime
as a toy model. Since we are interested primarily in the intertwining of 
general covariance with quantum field theory, the detailed dynamics  of the 
quantum field itself is a further complication which we may ignore in a 
first treatment.

Thus, we shall focus on the quantization of a free massive scalar field on arbitrary foliations of flat spacetime. An elegant way to view classical free scalar field theory on arbitrary foliations is to cast it in a generally covariant form known as parametrised field theory \cite{kkiyer}. In this form the theory can be used as testing ground for various aspects of general covariance encountered in gravity. Indeed, certain midisuperspace reductions of gravity such as
cylindrical waves \cite{kkcyl}, as well as theories of gravity in lower dimensions \cite{kkjoe} can be mapped onto parametrised field theory by suitable variable redefinitions, thus providing an even stronger motivation for studying 
parametrised field theory.

The canonical quantization of parametrised free field theory was studied
in  \cite{kk,ctmv1,ctmv2} 
 with interesting consequences such as the necessity of an anomaly potential
in the functional Schroedinger equation in 2 dimensions \cite{kk} and 
the non-existence of the functional Schroedinger picture as the unitary image of the Heisenberg picture in spacetime dimensions greater than 2 \cite{ctmv2}. 
Such results underline the importance of the study of parametrised field theory
both in itself and in its role as a toy model for {\em canonical} quantum gravity.
The next logical step is to 
to examine which, if any, aspects of the
 {\em path integral} approach to gravity may be better understood by an analysis of 
the path integral quantization of parametrised field theory.

As emphasized earlier, one of the problems of defining the  quantization of a 
generally covariant theory such as gravity is the absence of a preferred choice of time \cite{kktime}. This ``problem of time'' has been studied, most often,
in the canonical quantization context. In a path integral formulation
it is most directly encountered in the construction of vacuum wave functions.
In Poincare invariant theories, vacuum wave functions are constructed as
Euclidean path integrals which, in turn, are constructed   
from their Lorentzian counterparts by a Wick rotation of the preferred 
inertial time. In a generally covariant context no preferred time, and hence, no preferred Wick rotation is available to define the Euclidean theory.

Another aspect of the problem of time is that of inequivalent quantizations.
In canonical treatments of gravity the choice of time is very often made by
breaking the time reparameterization invariance of the theory via a choice of 
gauge fixing. Different choices of gauge fixing lead to different choices of 
time which in turn may lead to inequivalent quantizations. In  
a path integral for a theory with gauge invariances, gauge fixing terms must be included \cite{ggefix} so as to avoid infinities coming from summing over 
gauge equivalent configurations. Thus we expect that the problem of time could manifest in different choices of gauge fixings in the gravitational path 
integral. \footnote{ The Fadeev-Popov determinants in the path integral are supposed to ensure gauge independence. Note, however, that
the formal proof of gauge independence assumes that the  reduced phase space 
path integral implements a unique quantization. In theories without extra structures such as global Poincare invariance or in quantizations which do not
assign an explicit role to Poincare invariance, it is by no means clear that there
exists  a 
unique quantization at the reduced phase space level. 
Thus,
 a particular gauge choice may present the theory in a guise which is 
amenable to a particular choice of quantization.}

In this work we examine the above facets of the problem of time in the context of a path integral quantization of parametrised field theory. 
In section 2 we construct the (Lorentzian)
 configuration space path integral from the 
phase space path integral and hence obtain the correct non-trivial measure.
It is clearly seen that different time slicings correspond to appropriately 
different choices of gauge fixing terms. In section 3, we examine the 
issue of Euclideanization. As mentioned earlier, in the generally covariant 
context of gravity, no preferred Wick rotation to a Euclidean theory is 
available. Instead {\em ad hoc} prescriptions have been proposed 
\cite{hawkingeuc} which have no clear connection to the Lorentzian theory.
An exception is the proposal of Schleich \cite{kristin}, wherein the vacuum
wave function is defined from a reduced phase space path integral. We use her
ideas to define `Euclidean' path integrals in the generally covariant 
context of parametrised field theory. We find that her unambiguous definition
of Euclideanization implies the existence of non-standard Wick rotations of the standard free scalar field 2 point function. We confirm by direct inspection
that such Wick rotated 2 point functions indeed exist.

We initiate our investigation into the existence (or absence) of 
inequivalent quantizations for non- standard choices of time
in section 4. We show how
computations of the scalar field 2 point function may be used to illuminate
this issue. We work through, in some detail, the case of a time independent
foliation in 1+1  dimensions. Since there is reason to expect that this simple 
choice of time reproduces the standard quantization
\footnote{Our choice of foliation is such that 
the orbits of the time vector field defined by the foliation agree with
the orbits of the time isometry of the flat spacetime metric.}, we 
provide explicit
calculations primarily to illustrate our general framework. Indeed, the case 
of a general foliation is still open. 
In section 5 we discuss our results in the context of the path integral 
approach to quantum gravity as well as in the context of canonical quantization
of parametrised field theory and indicate open issues. Details of some of our
considerations are collected in appendices A1 and A2.

\noindent{\em Notation}:The flat n+1 dimensional spacetime manifold  is $R^{n+1}$. 
$\alpha ,\beta ,\gamma = 0..n$ are spacetime indices in an arbitrary coordinate 
system $x^{\alpha}$. We shall set $x^0 =t$. The $t$=constant submanifolds are assumed to
be   $n$ dimensional, spatial hypersurfaces  
diffeomorphic to $R^n$. $i,j,k=1..n$ are spatial indices on this hypersurface.
$A,B,C=0..n$ are spacetime indices in inertial coordinates $X^A$ with $X^0=:T$. The spatial inertial coordinates are $X^{\hat A},{\hat A} =1..n $. The Minkowski metric of signature
(-, ++..+) is $\eta_{\alpha \beta}$. $\partial_{\alpha}, \partial_{A}, \partial_i$ are the
partial derivative operators with respect to  $x^{\alpha}, X^A, x^i$ respectively. The dot 
`$\;\;\dot{}\;\;$'  denotes $\partial \over \partial t$.

\section*{2. The path integral}
In this section we derive the classical phase space action, define the phase space path integral and 
integrate over the momenta to obtain the configuration space path integral.

\subsection*{2.1 The classical formulation}
The Minkowski metric in a arbitrary coordinate system is given by 
\begin{equation}
\eta_{\alpha \beta} = \eta_{AB} \partial_{\alpha}X^A \partial_{\beta}X^B
\label{eta}
\end{equation} 
where $\eta_{AB}$ is the standard Minkowski metric in inertial coordinates.
From (\ref{eta}) the spacetime element in an arbitrary coordinate system is
\begin{equation}
ds^2 = (-{\dot T}^2 + {\dot X}^2)dt^2 + 
2(-{\dot T}\partial_i T+ {\dot {X^{\hat A}}}\partial_i X^{\hat A})dt dx^i
+ ( \partial_i X^{\hat A} \partial_j X_{\hat A} -  \partial_i T    \partial_j T) dx^i dx^j   . 
\label{ds2}
\end{equation} 
The line element may also be written in the standard ADM form in terms of the lapse $N$, shift $N^i$ and
spatial metric $q_{ij}$ as
\begin{equation}
ds^2 = (-N^2 + N_iN^i)dt^2 + 
2N_idt dx^i
+ q_{ij}dx^i dx^j   . 
\label{ds2adm}
\end{equation}
$X^A_{i}$ are the projectors into the hypersurface and are defined by 
\begin{equation}
X^A_{i} = \partial_i X^A .
\end{equation}
It is straightforward to show the following useful identities.
\begin{equation}
X^A_i X_{Aj} = q_{ij} .
\label{qij}
\end{equation}
\begin{equation}
\epsilon_{A A_1..A_n} X^{A_1}_{i_1}..X^{A_n}_{i_n}  = -n_A \epsilon_{i_1 i_2....i_n}
\label{epsilon}
\end{equation}
where $\epsilon_{A A_1..A_n}$ is the spacetime volume form, $\epsilon_{i_1 i_2....i_n}$ is the spatial
volume form on the $t$=constant spatial hypersurface and $n_A$ is the unit, future-pointing, timelike normal to this 
hypersurface. From (\ref{epsilon}) and $\sqrt{\eta} = N\sqrt{q}$ ($\eta,q$ are, respectively, 
the determinants of the spacetime and spatial metrics) it follows that
\begin{equation}
{\partial N\over \partial {\dot X}^A} = -n_A .
\label{nxdot}
\end{equation}
Equations (\ref{ds2}) and (\ref{ds2adm}) imply that
\begin{equation}
{\partial N^i\over \partial {\dot X}^A} =q^{ij}X_{Ai} .
\label{nixdot}
\end{equation}

The action for a free scalar field $\phi$ of mass $m$ on Minkowski spacetime expressed in inertial coordinates
 is 
\begin{equation}
S[\phi ]= -{1\over 2}\int d^{n+1}X 
(\eta^{AB}\partial_A\phi \partial_B\phi + m^2 \phi^2).
\label{action}
\end{equation}
The action for parametrised field theory is obtained by expressing  (\ref{action})  in 
arbitrary coordinates $x^{\alpha}$ and treating the action as a functional of $\phi$ as well as 
the embedding variables $X^A(x^i,t)$ . Thus
\begin{equation}
S[\phi, X^A] =-{1\over 2}
 \int d^{n+1}x{\sqrt{\eta}}(\eta^{\alpha \beta}\partial_{\alpha}\phi 
\partial_{\beta}\phi + m^2\phi^2) ,
\label{actionpft}
\end{equation}
with $\eta_{\alpha \beta}$ interpreted as a functional of $X^A$ via (\ref{eta}). 
In this form, the action is a manifestly diffoemorphism invariant functional of the 
$n+1$ scalar fields $X^A$ and the scalar field $\phi$.
A straightforward Hamiltonian analysis of (\ref{actionpft}) using (\ref{eta})-(\ref{nixdot}) yields the 
Hamiltonian form of the action given by 
\begin{equation}
S= \int dt d^nx (P_A {\dot X}^A  + \pi {\dot \phi} - M^A C_A) .
\label{hamaction}
\end{equation}
Here $P_A$ and $\pi$ are the momenta canonically conjugate to $X^A$ and $\phi$.
$M^A$ are the Lagrange multipliers for the first class constraints $C_A$ with 
\begin{equation}
C_A = P_A - n_A h + q^{ij}X_{Aj} h_i ,
\label{ca}
\end{equation}
\begin{equation}
h := {\pi^2 \over 2 {\sqrt q}} +{{\sqrt q}
(q^{ij}\partial_i \phi \partial_j \phi + m^2\phi^2)\over 2}  ,
\;\;\;\;\; h_i = \pi \partial_i \phi .
\label{hhi}
\end{equation}
Note that $n_A$ and $q_{ij}$ are to be considered as functionals of $X^A$ through (\ref{qij}) and
(\ref{epsilon}). The constraint algebra is abelian i.e. $\{C_A, C_B \}=0$. The algebra of diffeomorphisms
can be recovered by smearing the constraints with  vector fields $\xi_1, \xi_2$ which depend on the
embedding variables $X^A$ so that
\begin{equation}
\{ \int d^nx_1 \xi_1^AC_A , \int d^nx_2\xi_2^BC_B \} = \int d^nx (\xi_2^B\partial_B\xi_1^A - 
                                                   \xi_1^B\partial_B\xi_2^A)C_A      .   
\end{equation}
We restrict attention to asymptotically inertial embeddings by imposing the following 
boundary conditions as $\sum_{i=1}^n x^ix^i \rightarrow \infty$:
\begin{eqnarray}
X^0(x,t) = t, &  X^1(x,t)= x^1,\;\; X^2(x,t)= x^2,...,\;X^n(x,t)= x^n,  \label{bdryx}\\ 
M^0 (x,t)= 1,  & M^{\hat A} = 0, \;\; {\hat A} =1,..,n . \label{bdrym}\\
\end{eqnarray}
We also impose that $P_A, \pi , \phi$ be of compact support on the spatial slice. 

\subsection*{2.2 The Path Integral}

In addition to the classical action (\ref{hamaction}), a choice of gauge fixing is needed to define the phase space path integral. Since this work constitutes
a first attempt to analyse the problem of time in a path integral context
in parametrised field theory, we restrict attention to choices of gauge fixing 
which have the clear geometric meaning of  fixing corresponding choices
 of time functions (i.e. foliations by spacelike surfaces of constant time)
on the flat spacetime. The gauge fixing term $\delta [ \chi^A ]$
where 
\begin{equation}
\chi^A = X^A(x,t) - f^A(x,t), 
\label{gf}
\end{equation}
corresponds to choosing a foliation of the spacetime defined by 
the embedding variables $X^A(x,t)$ taking the values $f^A(x,t)$.
With this choice of gauge fixing it is easily checked that the Fadeev-Popov
determinant (see for example \cite{sundermeyer}) is unity. Hence, the phase space
path integral is given by 
\begin{equation}
Z = \int {\cal D} \phi{\cal D}\pi {\cal D}M^A{\cal D}P_A{\cal D}X^A
      \delta [X^A= f^A]
    \exp i\int dtd^nx (P_A {\dot X}^A +\pi {\dot {\phi}} - M^AC_A).
\label{hampi}
\end{equation}
In the above equation it is understood that the configuration variables
$\phi (x,t), X^A (x,t)$ interpolate between between fixed initial and final 
 values at some initial and final instants of time $t=t_I$ and $t=t_F$.
The end point values of $X^A$ are assumed to be 
consistent with the gauge choice (\ref{gf}). We shall not explicitly 
specify the end point dependence of $Z$ in our notation. 

We integrate (\ref{hampi}) over $M^A$ and $P_A$ to obtain
\begin{equation}
Z = \int {\cal D} \phi{\cal D}\pi {\cal D}X^A
      \delta [X^A= f^A]
    \exp i\int dtd^nx (\pi {\dot{\phi}} -Nh -N^ih_i)
\label{hampi2}
\end{equation}
with 
\begin{equation}
N= -{\dot X}^A n_A,  \;\;\; N^i= q^{ij}X_{Ai}{\dot X}^A
\label{nni}
\end{equation}
(notice the consistency of these expressions 
with (\ref{nxdot}), (\ref{nixdot})),
and $h,h_i$ given by (\ref{hhi}).
In this form it is clear that our choice of gauge fixing presents the parametrised field theory in the form of free scalar field theory on the (in 
general, curved,) foliation $f^A(x,t)$.
A further integration over the momenta, $\pi$, yields the configuration space path integral
\begin{equation}
Z = \int {\cal D} \phi {\cal D}X^A [{\rm det}{iN\over \sqrt{q}}]^{-{1\over 2}} 
      \delta [X^A= f^A]
    \exp iS[\phi ,X^A]
\label{lpi}
\end{equation} 
where $S[\phi ,X^A]$ is the classical action given by (\ref{actionpft}).

Note that the path integral measure has a factor of 
$[{\rm det}{iN\over \sqrt{q}}]^{-{1\over 2}}$. In our specific gauge choice,
this determinant factor reduces to an irrelevant c- number depending 
on $f^A(x,t)$. However, with a  more general choice of gauge fixing term,
we expect this term to persist and, as a consequence, contribute non-trivially
to the path integral measure. Although the treatment of the most general gauge
choice can be done via BRST methods (see \cite{fv2,kristin}), such a treatment
is beyond the scope of this paper. Instead, in the appendix, we have extended
our treatment to slightly more 
general ($\phi$ - dependent) gauge choices than those of (\ref{gf}).
We find that the determinant factor persists and contributes non-trivially
to the measure. We believe that this measure is the exact analog of the 
measure found by Fradkin and Vilkovisky in  \cite{fv} for quantum gravity.
This lends added credence to their measure being the correct one rather
than the more commonly used measure proposed by de Witt in \cite{dewitt}.
We shall comment further on this in section 6.

\section*{3. Euclideanization.}

Our aim is to construct convergent path integrals in order to 
evaluate vacuum wave functions. Indeed, we shall {\em define}  
this  construction to be Euclideanization. For the case 
of a flat inertial foliation, it will be seen that 
 the construction  reproduces the 
standard Wick rotated path integral.

We are
motivated by the remark of Schleich in \cite{kristin} to the effect that
there is no obstruction to 
constructing a
convergent path integral for the vacuum wave function in any theory
with a positive definite Hamiltonian. To illustrate this remark, consider
such a theory in the absence of constraints with a time independent
Hamiltonian, $H$, and a
single configuration space degree of freedom, $q$.
The vacuum is defined as the eigenfunction of ${\hat H}$ 
with lowest eigen value.
Under the assumption that the vacuum is unique and that the zero of energy has
been chosen so that the vacuum energy vanishes, the vacuum wave function 
(in obvious notation)
may be obtained from the Feynman-Kac type formula:
\begin{equation}
\psi_0 (q_F,t_F) \psi_0^*(q_I,t_I) = \lim_{t_I \rightarrow -\infty}
         <q_F, t_F| \exp (-ia{\hat H}(t_F-t_I)| q_I,t_I> .
\label{fkac}
\end{equation}
Here $a$ is any complex number with negative imaginary part and $t_F,t_I$ are
final and initial times. The above identity
is obtained by  by expanding $|q_F,t_F>,|q_I,t_I>$ in a complete set of 
energy eigen states. The negative imaginary part of $a$ and the 
$t_I\rightarrow -\infty$ limit conspire to project the initial and 
final 
states onto the vacuum state, resulting in equation (\ref{fkac}).
It is straightforward to check that 
the matrix element on the right hand side of this identity may be written
as a phase space path integral:
\begin{equation}
<q_F, t_F| \exp (-ia{\hat H}(t_F-t_I)| q_I,t_I>
= \int {\cal D}q{\cal D}p 
\exp (i\int dt (p_i {\dot q_i} - (a+1)H)).
\label{posham}
\end{equation}
Since $H$ is positive and $a$ has negative imaginary part, the path integral
is (formally) convergent.
Equations (\ref{fkac}) and (\ref{posham}) illustrate Schleich's
remarks in the context of systems without constraints.

In appendix A2, we 
develop   similar expressions for vacuum wave functions in terms of phase space
path integrals for systems with first class constraints. We restrict 
attention to cases in which time evolution is
generated by  a non- vanishing positive
definite Hamiltonian. 
In section 3.1,  we extend the  considerations of appendix A2 to parametrised 
field theory so as to write vacuum functions as convergent phase space path 
integrals.
In section 3.2, we integrate these expressions over momenta
to obtain convergent path integrals which we refer to as Euclidean 
path integrals for want of a better name. The results of section 3.2
imply that the standard Minkowskian  2 point function can be 
continued to a `Euclidean' 2 point function through non-standard
Wick rotations. We show that this is indeed true in section 3.3.

\subsection*{3.1 Vacuum wave functions as phase space path integrals.}

In a theory with coordinates $q_i$, momenta $p_i$ ( $i=1..n$),
first class constraints
$C_{\alpha}, \alpha=1..m$, 
Lagrange multipliers $\lambda^{\alpha}$, gauge fixing constraints
$\chi_{\alpha}$ and Hamiltonian $H$, the transition amplitude can be written as
\begin{equation}
Z(q_{iI}, t_I,q_{iF}, t_F )
= \int {\cal D}q{\cal D}p{\cal D}\lambda \delta (\chi_{\alpha})
{\rm det}[\{C_{\alpha}, \chi_{\beta} \}]
\exp (i\int p_i {\dot q_i} - \lambda^{\alpha}C_{\alpha} - H).
\label{pqamp}
\end{equation}
The integral  is over all paths which have endpoints at times $t_I, t_F$
specified
by initial values $q_i =q_{iI}$ and final values $q_i =q_{iF}$. The endpoint
configurations must satisfy the gauge fixing constraints.

Standard arguments \cite{sundermeyer} using canonical transformations to 
appropriate variables indicate that the transition amplitude (\ref{pqamp})
is independent of the gauge choice $\chi_{\alpha}$. These
arguments are not without shortcomings. First, any application of 
canonical transformations to the path integral is fraught with problems
related to the `roughness of paths' \cite{schulman}. Second, many of the
standard arguments (\cite{ht} is an exception) ignore endpoint contributions
in the canonically transformed action,
as well as the possible differences in the specification of endpoint values
in terms of the old configuration variables as compared
to their specification in terms of the new 
canonically transformed  configuration variables.

Here (and in the appendix A2), we shall 
also ignore the issues mentioned above. 
Although we shall further justify our constructions for parametrised 
field theory, these issues require a careful treatment in more 
complicated systems such as quantum gravity.
With these caveats in mind, 
we consider the gauge independent quantity $Z_a$ given by 
\begin{equation}
Z_{a}(q_{iI}, t_I;q_{iF}, t_F )
= \int {\cal D}q{\cal D}p{\cal D}\lambda \delta (\chi_{\alpha})
{\rm det}[\{ C_{\alpha}, \chi_{\beta} \}]
\exp (i\int dt (p_i {\dot q_i} - \lambda^{\alpha}C_{\alpha} - aH)),
\label{pqampa}
\end{equation}
where $a$ is an arbitrary complex number. 
As shown in the appendix, if we choose $a$ to have negative imaginary part
and the vacuum to have vanishing energy, 
it follows that
\begin{equation}
Z_{a}
(q_{iI}, t_I=-\infty ;q_{iF}, t_F )= \psi_0 (q_F,t_F) \psi_0^*(q_I,t_I),
\label{25}
\end{equation}
where $\psi_0$ denotes the vacuum wave function. For a fixed initial
configuration, this is an expression for the vacuum wave function
as a function of the final configuration.

We aim to construct similar phase space path integrals to express the 
vacuum wave functionals of parametrised field theory. Our strategy
is to first construct the counterpart of equation (\ref{pqampa})
and then to argue that its $t_I\rightarrow -\infty$
limit is the correct parametrised field theory counterpart of equation
 (\ref{25}). There are two differences between parametrised field theory and
the system considered in equation (\ref{pqampa}). First, 
the action for parametrised field theory given by
equation (\ref{hamaction})
does not exhibit a non-vanishing Hamiltonian and second, 
the gauge fixing conditions which define a foliation are {\em time dependent}
\footnote{ Note that the conditions $X^A(x,t) = f^A (x,t)$ consitute 
a {\em 1 parameter family} of gauge fixing conditions i.e. for every instant
of time one has a complete gauge fixing. Hence, strictly speaking, these
conditions define a {\em deparametrization} of the theory rather than a 
gauge fixing. We shall, however, continue to refer to them as gauge fixing 
conditions.}. 
Our considerations for the system defined by 
(\ref{pqampa}) generalise straightforwardly to parametrised field theory in 
spite of this time dependence. The first point of difference
remains, however, and we need to construct an 
analog of the Hamiltonian in (\ref{pqampa}). 
To do so, we note that in standard free scalar field theory, the vacuum is 
defined as the ground state of the operator corresponding to the conserved,
positive definite
scalar field energy `$E$' given by  
\begin{equation}
E= {1\over 2}
\int_{T={\rm constant}}d^{n}X( ({\partial \phi\over \partial T})^2 + 
                 \sum_{\hat A}\partial_{\hat A}\phi 
                              \partial_{\hat A}\phi  +m^2 \phi^2).
\end{equation}
In the context of parametrised field theory it is straightforward to verify
 that $E$ is simply the evaluation, on a classical solution, 
of the  Dirac observable,
$H$, given by 
\begin{equation}
H= \int_{R^n} d^{n}x P_T ,
\label{defh}
\end{equation}
where $P_T:=P_{A=0}$.
Therefore, we define the vacuum state in parametrised field theory to be the
ground state of the operator corresponding to $H$ (see (\ref{defh}).
Since $E=H$ on the constraint surface, our definition is consistent with 
the usual definition of the vacuum in free scalar field theory.

%
The above considerations  imply that the correct counterpart 
of
(\ref{pqampa}) is 
\begin{eqnarray}
Z_a[\phi_I,X^A_I, t_I; \phi_F, X^A_F,t_F] := 
\int {\cal D} \phi{\cal D}\pi {\cal D}M^A{\cal D}P_A{\cal D}X^A
      \delta [X^A= f^A] \nonumber\\
   \exp (i\int dtd^nx (P_A {\dot X}^A +\pi {\dot {\phi}} - M^AC_A) -
                            ia\int dt H ),
\label{hampia}
\end{eqnarray}
with $H$ defined by (\ref{defh}).
Using the fact that $H$ is a Dirac observable,
it can be checked that 
the methods of \cite{sundermeyer} mentioned in the appendix A2 show 
gauge independence 
(with respect to appropriately defined infinitesmal changes of gauge 
\cite{sundermeyer}) 
of the above expression. 

To see that the $t_I\rightarrow -\infty$ limit of (\ref{hampia}) indeed 
yields the vacuum wave function, we note that   
the integration of equation (\ref{hampia}) over $X^A$ and $M^A$ gives
\begin{equation}
Z_a = \int {\cal D} \phi{\cal D}\pi
    \exp i\int dtd^nx (\pi {\dot{\phi}} -{\cal H} -aH).
\label{calh}
\end{equation}
Here ${\cal H} = {\dot F}^Ah_A$ is the generator of evolution in time $`t$' 
along  the foliation $F^A(x^i, t)$. Therefore, in operator language
(with a suitable operator ordering prescription) the above path integral 
expression corresponds to, in obvious notation, 
\begin{eqnarray}
Z_a &=&
\lim_{t_I\rightarrow -\infty}< \phi_F, t_F |e^{-ia{\hat H}(t_F-t_I)} 
| \phi_I, t_I> \\
&=& \psi_0[\phi_F ;t_F]\psi^*_0[\phi_I ;t_I],
\label{zevac}
\end{eqnarray}
where $\psi_0$ denotes the vacuum state and we have evaluated the action 
of $e^{-ia{\hat H}(t_F-t_I)}$ via a spectral decomposition of 
${\hat H}$ under the assumption that  its lowest eigen value is normalised
to zero and that $a$ has negative imaginary part.
Equation (\ref{zevac}) further justifies our constructions for 
Euclideanization in parametrised field theory.

Finally, we note that on the 
flat foliation,
$f^A= x^{\alpha} \delta_{\alpha}^A$, $Z_a$ evaluates to
\begin{equation}
Z_{a}=\int {\cal D} \phi{\cal D}\pi   \exp i\int dtd^nx (\pi {\dot {\phi}} -
                             {a+1\over 2 }(\pi^2 + 
     \sum_{i}\partial_{i}\phi 
                              \partial_{i}\phi  +m^2 \phi^2) ).
\label{a1i}
\end{equation}
The choice $a=-1-i$ reproduces the usual 
expression for the vacuum wave functional.
Henceforth we shall set $a=-1-i$ and define the Euclidean phase space
path integral to be 
\begin{eqnarray}
Z_{E}[\phi_I,X^A_I, t_I=-\infty; \phi_F, X^A_F,t_F] 
= \int {\cal D} \phi{\cal D}\pi {\cal D}M^A{\cal D}P_A{\cal D}X^A
      \delta [X^A= f^A] \nonumber\\
\exp i\int dtd^nx (P_A {\dot X}^A +\pi {\dot {\phi}} - M^AC_A -
                            (-1-i)P_T ) . 
\label{defze}
\end{eqnarray}

Due to the positivity of (\ref{defh}) on the constraint surface,
the above expression is (formally) convergent. Our strategy is to 
define the ``Euclidean'' theory in configuration space by integrating
over the momenta in (\ref{defze}).

\subsection*{3.2 Euclidean path integral.}

Equation (\ref{defze}) can be written in the form
\begin{eqnarray}
Z_{E}[\phi_I,X^A_I, t_I=-\infty; \phi_F, X^A_F,t_F] 
= \int {\cal D} \phi{\cal D}\pi {\cal D}M^A{\cal D}P_A{\cal D}X^A
      \delta [X^A= f^A] \nonumber\\
\exp i\int dtd^nx (P_{T} {\dot T}_E +
P_{\hat A} {\dot X}^{\hat A} +\pi {\dot {\phi}} - M^AC_A 
                             ) 
\label{defzte}
\end{eqnarray}
where we have defined 
\begin{equation}
T_E = T - t - it .
\label{defte}
\end{equation}
Integration over $M^A, P_A$ yields
\begin{equation}
Z_E = \int {\cal D} \phi{\cal D}\pi {\cal D}X^A
      \delta [X^A= f^A]
    \exp i\int dtd^nx (\pi {\dot{\phi}} -N_Eh -N_E^ih_i)
\label{zene0}
\end{equation}
where $h,h_i$ are given by (\ref{hhi}) and we define
\begin{equation}
N_E= -{\dot X}^{A} n_{A} + (1+i)n_{A=0}, 
\label{ne}
\end{equation}
\begin{equation}
N^i_E= q^{ij}(X_{{A}j}{\dot X}^{A} -(1+i)\partial_jT).
\label{nie}
\end{equation}
Notice that the above equations can be obtained from equations 
(\ref{hampi2}) and (\ref{nni}) by replacing $T$ with $T_E$.

Next, we integrate over $\pi$. After ``completing the square'',
the $\pi$- dependent term to be integrated over is
\[
\exp -i {N_E \sqrt{q} \over 2} 
(\pi - {{\dot \phi}- N^i_E\partial_i \phi \over N_E})^2 .
\]

In what follows, we shall denote the real and imaginary parts of a complex
number $a$ by $a_R$ and $a_I$ so that $a = a_R +i a_I$. The absolute value
of $a$ will be denoted by $|a|$. 
From (\ref{ne}) and from the fact that $n^A$ is a future pointing,
timelike vector, we have that
\begin{equation}
N_{E_I} = n_{A=0} = n_A ({\partial \over \partial T})^A <0 .
\label{neineg}
\end{equation}
This ensures that the exponential is convergent and can be integrated
over $\pi$. Performing this integration yields
\begin{equation}
Z_E = \int {\cal D} \phi{\cal D}X^A
      \delta [X^A= f^A]
      \exp -\int d^{n+1}x L_E, 
\label{zene}
\end{equation}
where $L_E$ is given by
\begin{equation}
L_E= -{i\over 2} N_E{\sqrt q} 
(({{\dot \phi}- N^i_E\partial_i \phi \over N_E})^2 - q^{ij}\partial_i\phi
            \partial_j \phi -m^2\phi^2).
\label{le}
\end{equation}
This is our final expression for what we call the Euclidean path integral.

Next, we show that the Euclidean path integral is indeed convergent by
showing that the real part of $L_E$ is positive.
From (\ref{le}), the real part of $L_E$ is determined by the expression
\begin{equation}
-{2L_{E_R} \over \sqrt{q}} =
N_{E_I}(q^{ij} - {N^{i}_{E_I}N^{j}_{E_I}\over |N_E|^2}) A_i A_j
+ {N_{E_I}\over |N_E|^2}B^2 + 2{N^{i}_{E_I}N_{E_R}\over |N_E|^2} A_iB
+N_{E_I}m^2\phi^2,
\label{lr}
\end{equation}
where
\begin{equation}
A_i = \partial_i \phi \;\;\;\;\;\;
B= {\dot \phi}- N^i_{E_R}\partial_i \phi
\label{defab}
\end{equation}

It is straightforward to show, using equations (\ref{nni}),(\ref{qij}), 
(\ref{ne}) and 
(\ref{nie}), that
\begin{equation}
N_{E_I}^2 - q_{ij} N^i_{E_I}N^{j}_{E_I} = 
{1\over 2} {\partial^2\over \partial {\dot T}^2}(N^2 - q_{ij}N^iN^j) .
\end{equation}
Using  this in conjunction with equation (\ref{ds2adm}) gives the 
key inequality, 
\begin{equation}
N_{E_I}^2 - q_{ij} N^i_{E_I}N^{j}_{E_I} =
{1\over 2} {\partial^2\over \partial {\dot T}^2} ({\dot T}^2 -{\dot X}^2)=1>0.
\label{keyineq}
\end{equation}
Note that a trivial application of the Schwarz inequality shows that 
\begin{equation}
N_{E_I}^2 q^{ij} A_i A_j > (N^{i}_{E_I} A_i)^2 .
\label{schwarz}
\end{equation}

Straightforward manipulations using the above inequalities in conjunction with 
(\ref{neineg}) imply that, when $A_i$ and $B$ are not both identically
zero,
\begin{equation}
-{2L_{E_R}\over \sqrt{q}} < 
{N_{E_I}\over |N_E|^2}(|B| -|N_{E_R}|\sqrt{q^{ij}A_i A_j})^2 +
N_{E_I}m^2\phi^2 < 0 .
\end{equation}
Thus, as expected, the Euclidean path integral is convergent.

For a flat foliation with $T=t$, equation(\ref{defte}) defines
a Euclidean time via the standard Wick rotation. Note that we did {\em not} 
first define a Lorentzian configuration space path integral and then make a Wick rotation. Rather, we defined a `Euclidean' phase space integral and 
equation (\ref{defte}) emerged as a consequence of this. For arbitrary foliations, we have that $t\neq T$ and consequently that  equation (\ref{defte}) {\em differs} from the standard Wick rotation!
This suggests that the 2 point functions of the theory can be continued
through this non-standard Wick rotation. We shall confirm the
existence of these Wick rotated 2 point functions in section 3.3.
The Euclidean action (\ref{le}) is in general complex and depends
on the choice of foliation as does the Wick rotation. 
This is reminiscent of `t Hooft's discussion
of Wick rotations in perturbative quantum gravity \cite{thooft} wherein
he states that
the details of the Wick rotation depend on the gauge chosen.

\subsection*{3.3 `Wick rotated' 2 point functions.}

In order to discuss Wick rotations of the form (\ref{defte}), it is useful
to express the embedding time $X^0= T$ in terms of $x^{\alpha}$ through
a function $h (x^{\alpha})$ defined by 
\begin{equation}
T= t + h (x^{\alpha}).
\label{defhalpha}
\end{equation}
From (\ref{defte}), the `Euclidean' time $T_E$ is 
\begin{equation}
T_E = -it +h (x^{\alpha}).
\label{defhE}
\end{equation} 

Denote the standard time- ordered Minkowski spacetime  2 point function by 
$G(T_1, X^{\hat A}_1; T_2, X^{\hat A}_2)$. $G$ is a function of 
$x^{\alpha}_i, i=1,2$ through the dependence on $x^{\alpha}$ of the embeddings
 i.e. $X^A \equiv X^A(x^{\alpha})$. The Wick rotated 2 point function
$G_E$ is given by 
\begin{equation}
G_E(x^{\alpha}_1, x^{\alpha}_2)
= G (T_{1E}(x^{\alpha}_1), X^{\hat A}_1(x^{\alpha}_1);
T_{2E}(x^{\alpha}_2), X^{\hat A}_2(x^{\alpha}_2)),
\label{gE}
\end{equation}
where the right hand side denotes a continuation of $G$ to 
the complex arguments defined by (\ref{defhE}).

To show that $G_E$ exists, recall that 
the standard Minkowsian 2 point function is defined by 
\begin{equation}
G(X^A_1, X^A_2) =
<0| \theta (T_1- T_2) {\hat \phi} (X^A_1) {\hat \phi} (X^A_2)
      +\theta (T_2- T_1) {\hat \phi} (X^A_2) {\hat \phi} (X^A_1)|0>
\end{equation}
where $|0>$ denotes the vacuum state. As shown in section 4 (see the 
discussion after equation(\ref{o}) ), due to Lorentz invariance and the 
spacelike nature of $t=$ constant slices, we can equally well write the 
2 point function as
\begin{equation}
<0|\theta (t_1- t_2) {\hat \phi} (X^A_1(x^{\alpha}_1)) 
{\hat \phi} (X^A_2(x^{\alpha}_2))
      +\theta (t_2- t_1) {\hat \phi} (X^A_2(x^{\alpha}_2)) 
{\hat \phi} (X^A_1(x^{\alpha}_1))|0>.
\end{equation}
Denoting $<0|{\hat \phi} (X^A_1(x^{\alpha}_1)) 
{\hat \phi} (X^A_2(x^{\alpha}_2))|0>$ by 
$D(X^A_1(x^{\alpha}_1), X^A_2(x^{\alpha}_2))$ 
we can write the 
above equation as 
\begin{eqnarray}
G(X^A_1(x^{\alpha}_1),X^A_2( x^{\alpha}_2))|_{t_1>t_2}
&= & D(X^A_1(x^{\alpha}_1), X^A_2(x^{\alpha}_2))|_{t_1>t_2} \\
G(X^A_1(x^{\alpha}_1),X^A_2( x^{\alpha}_2))|_{t_2>t_1}
& =& D(X^A_2(x^{\alpha}_2), X^A_1(x^{\alpha}_1))|_{t_2>t_1} \\
G(X^A_1(x^{\alpha}_1), X^A_2(x^{\alpha}_2))|_{t_1=t_2}
& =& D(X^A_1(x^{\alpha}_1), X^A_2(x^{\alpha}_2))|_{t_1=t_2} \\
& =& D(X^A_2(x^{\alpha}_2), X^A_1(x^{\alpha}_1))|_{t_1=t_2}. 
\end{eqnarray}
To obtain the last equation, note that $t_2=t_1$ implies that the events
1 and 2 are spacelike and hence the field operators at these points commute.
We use the standard expression for $D(X^A_1, X^A_2)$ 
and equation (\ref{defhalpha}) to obtain
\begin{eqnarray}
D(X^A_1, X^A_2) & = & 
({1\over 2\pi})^n \int {d^nk\over 2\omega_k} 
e^{-i\omega_k (T_1-T_2) + i k_i(X^i_1- X^i_2)}
\nonumber\\
&=&
({1\over 2\pi})^n \int {d^nk\over 2\omega_k} 
e^{-i\omega_k (t_1-t_2) +
i (k_i(X^i_1(x^{\alpha}_1)- X^i_2(x^{\alpha}_2))- \omega_k (h(x^{\alpha}_1)
                               -h(x^{\alpha}_2))},
\end{eqnarray}
where $\omega_k = \sqrt{\sum_{i=1}^n (k_i)^2 + m^2}$.
From (\ref{gE}) and the above equations we obtain
\begin{eqnarray}
G_E(x^{\alpha}_1, x^{\alpha}_2)|_{t_1>t_2}
&=&
({1\over 2\pi})^n \int {d^nk\over 2\omega_k} 
e^{-\omega_k (t_1-t_2) + i (k_i(X^i_1(x^{\alpha}_1)- X^i_2(x^{\alpha}_2))- \omega_k (h(x^{\alpha}_1)
                               -h(x^{\alpha}_2)))},\\
G_E(x^{\alpha}_1, x^{\alpha}_2)|_{t_2>t_1}
&=&
({1\over 2\pi})^n \int {d^nk\over 2\omega_k} 
e^{-\omega_k (t_2-t_1) + i (k_i(X^i_2(x^{\alpha}_2)- X^i_1(x^{\alpha}_1))- \omega_k (h(x^{\alpha}_2)
                               -h(x^{\alpha}_1)))},\\
G_E(x^{\alpha}_1, x^{\alpha}_2)|_{t_1=t_2}
&=&
({1\over 2\pi})^n \int {d^nk\over 2\omega_k} 
e^{i (k_i(X^i_1(x^{\alpha}_1)- X^i_2(x^{\alpha}_2))- \omega_k (h(x^{\alpha}_1)
                               -h(x^{\alpha}_2)))}\\
&=&
({1\over 2\pi})^n \int {d^nk\over 2\omega_k} 
e^{  i (k_i(X^i_2(x^{\alpha}_2)- X^i_1(x^{\alpha}_1))- \omega_k (h(x^{\alpha}_2)
                               -h(x^{\alpha}_1)))}.
\end{eqnarray}
Clearly the first two equations above define convergent integrals whereas
the expressions for $t_1=t_2$ agree with their Lorentzian counterparts
and hence exist as well defined distributions.

Thus, direct evaluation of (\ref{gE}) indeed shows that the Wick rotated 2 point functions do exist! Note that in the generally covariant formulation of 
parametrised field theory, the coordinate $t$, which is crucial to the 
definition of the Wick rotation (\ref{defhE}), has no intrinsically distinguished
role. Indeed, when confronted with parametrised field theory, it is 
difficult to guess the existence of this foliation- dependent Wick rotation 
and the consequent Euclideanization of the theory. From this perspective it
is very satisfying  to see that the well motivated definition of
Euclideanization which we have used is in harmony with the properties of the
standard Minkowskian 2 point function.

\section*{4. Path integral quantization with non-standard choice of time.}

The form of the scalar field action appropriate to the $f^A$ foliation may be 
obtained either by integrating the path integral (\ref{lpi}) over the 
embedding variables or by performing a coordinate transformation from
inertial coordinates $X^A$ to $x^{\alpha}$ in the action (\ref{action}).
The action describes  a scalar field on a flat Minkowski spacetime
{\em with inertial coordinates $x^{\alpha}$} interacting with an external
field determined by $f^A(x,t)$. The issue of interest is whether, despite being classically
equivalent to the standard action (\ref{action}), the action in this form naturally suggests a quantization procedure based on the $x^{\alpha}$- flat spacetime
which is inequivalent to the standard quantization. Since the action is 
quadratic in the scalar field, most of the physics is in the  2 point function.
An application of 
standard perturbative quantum field theory to this form of the action
results in a computation of the 2 point function in an expansion in powers
of the external field. The result can be compared to the standard Minkowskian
2 point function.

Let $p_i, i=1,2$  be a pair of events on the flat spacetime. We denote their spacetime 
coordinates in the $x^{\alpha}$ coordinate system by $\xv_i$ and their coordinates in 
the inertial $X^A$ system by $\Xv_i$. Let the 2 point function in the $f^A$
formulation be $G_f(\xv_1, \xv_2)$ and let the standard 2 point function be
$G(\Xv_1, \Xv_2)$.
In the case of the non standard foliation,
the 2 point operator is
\begin{equation}
{\hat O}_f = \theta (t_1- t_2) {\hat \phi} (p_1) {\hat \phi} (p_2)
      +\theta (t_2- t_1) {\hat \phi} (p_2) {\hat \phi} (p_1), 
\label{of}
\end{equation}
whereas in the usual inertial foliation the 2 point operator is
\begin{equation}
{\hat O} = \theta (T_1- T_2) {\hat \phi} (p_1) {\hat \phi} (p_2)
      +\theta (T_2- T_1) {\hat \phi} (p_2) {\hat \phi} (p_1). 
\label{o}
\end{equation}
Here $\theta$ is the usual step function which implements time ordering.
It is readily verified, using the fact that the $t=$ constant slices are spacelike with future pointing timelike normal $(dt)_{\alpha}$, that if $p_1$ and
$p_2$ are causally related then they have the same time ordering with respect
to $t$ as with $T$. Further, in the standard quantization, if $p_1$ and
$p_2$ are not causally related then $[{\hat \phi}(p_1),{\hat \phi}(p_2)]$
vanishes and the ordering doesnt matter. Thus, we can as well replace 
${\hat O}$ by ${\hat O_f}$ in the standard quantization.

If $G_f (\xv_1, \xv_2)\neq G(\Xv_1, \Xv_2)$, we may conclude that either the 
$f$ dependent quantization and the standard quantization  of the operator ${\hat O}_f$ are inequivalent  or that the 
vacuum state selected by the procedure to calculate $G_f$ is different from the standard vacuum state. Thus, in both cases the choice of foliation affects the quantum theory either in its representation of operators or in its 
identification of the vacuum state. 
To illustrate our ideas, we work 
through a simple 1+1 dimensional example in section 4.1. In section 4.2 we 
describe our framework for a general foliation in $n+1$ dimensions.
The actual computations which would show existence (or lack thereof) of 
inequivalent quantizations are left for future work.

\subsection*{4.1 Two dimensional example.}

In 2 spacetime dimensions
denote the inertial time by $T$ and the inertial space coordinate by $X$
and specify the foliation by  
$T(x,t) = t+ f(x)$ and  $X(x,t) = x$, where $f(x)$ is a function of 
compact support.
The path integral (\ref{lpi}) can be integrated over $X^A$ to give
\begin{equation}
Z = \int {\cal D} \phi 
      \exp iS[\phi(x,t)].
\label{2dlpi}
\end{equation}
The irrelevant c-number determinant  has been dropped and 
$S[\phi(x,t)]$ is defined as
\begin{equation}
S[\phi (x,t)] ={1\over 2}
 \int dx dt (\eta^{\alpha \beta}\partial_{\alpha}\phi \partial_{\beta}\phi 
     + f^{\alpha \beta}\partial_{\alpha}\phi \partial_{\beta}\phi
-m^2\phi^2) .
\label{2daction}
\end{equation}
Here $\eta_{\alpha \beta}$ denotes the flat metric with line element
$ds^2 = (dt)^2 - (dx)^2$ and $f^{\alpha \beta}$ is defined as
\begin{equation}
f^{00} = -({df\over dx})^2,   \;\;\;\;\;\; f^{01}=f^{10}= {df\over dx}
\;\;\;\;\; f^{11}=0 .
\label{fdef}
\end{equation}
The reader is requested to bear with us, in that we have changed our 
conventions for the metric signature from (-+) to (+-)
{\em only in this subsection}. The reason is to ensure easy cross-checking of numerical factors for Feynman diagrams with standard field theory references 
(see, for example, \cite{iz}) which use the (+-) conventions.

The action (\ref{2dlpi}) describes a scalar field interacting with a static 
potential on the $(x,t)$- Minkowski spacetime and $G_f$ may be computed via
standard Feynman diagrammatics. In momentum space, we have
\begin{equation}
G_f (\pv ,\qv ) := \int d^2x d^2y e^{i\pv \cdot\xv}e^{-i\qv \cdot\yv}
G_f(\xv, \yv ).
\end{equation}
where we have used the notation $\xv$ for $(x^0, x)$, $\pv$ for $(p^0, p)$
and $\pv \cdot \xv$ for $(p^0x^0 - px)$.
The Fourier transform of $f^{\alpha \beta} (\xv)$ is defined as 
\begin{equation}
f^{\alpha \beta} (\kv )= \int d^2x f^{\alpha \beta} (\xv) e^{i\qv \cdot\xv} .
\end{equation}

No loops are encountered in the Feynman diagrams 
since (\ref{2daction}) has only 2 point interactions. 
Each vertex contributes a factor of
\begin{equation}
iC(\kv_1,\kv_2) = if^{\mu \nu}(\kv_1 -\kv_2)k_{1\mu} k_{2\nu}
\label{cdef}
\end{equation}
with incoming momentum $\kv_1$ and outgoing momentum $\kv_2$. Each propagator
contributes a factor of ${-i\over \kprop}$. Here $\kv^2=\kv\cdot\kv$.
Notice that the choice of time `$t$' dictates the $i\epsilon$ prescription
in the propagator.

With the correct factors of $i$ and $2\pi$ we have 
\begin{eqnarray}
G_f(\pv ,\qv ) & = & {i(2\pi )^2\over \pprop}\delta (\pv , \qv)
                 + {-i\over \pprop} C(\pv , \qv ){1\over \qprop} \nonumber\\
 & + & \sum_{n=2}^{\infty}{(-1)^n i\over (2\pi)^{2n-2}}{1\over \pprop}
\int \prod_{j=1}^{n-1}d^2k_j\big(C(\pv,\kv_1)
{1\over \kv_1^2-m^2 +i\epsilon}
\nonumber \\
 & &                 C(\kv_1,\kv_2)...C(\kv_{n-2},\kv_{n-1})
{1\over \kv_{n-1}^2-m^2 +i\epsilon}C(\kv_{n-1},\qv)\big) 
 {1\over \qprop} \label{gf2d}
\end{eqnarray}
From (\ref{fdef}) and (\ref{cdef}) we have that 
\begin{eqnarray}
C(\kv ,\lv) &= & C^{(1)}(\kv ,\lv) + C^{(2)}(\kv , \lv ), \\
C^{(1)}(\kv,\lv) &=& 2\pi i \delta (k_0,l_0) f(k-l) k_0 (k^2-l^2), 
\label{c1}\\
C^{(2)}(\kv,\lv) &=& k_0^2 \delta (k_0,l_0)
                \int ds f(s)f(k-l-s)s (k-l-s). 
\label{c2}
\end{eqnarray}
The $\delta (k_0,l_0)$  factors ensure that, as expected, the static potential
conserves energy. Of course, momentum is not conserved due to the lack of
translational invariance in the presence of the potential.
Using (\ref{c1}) and (\ref{c2}), we can write $G_f (\pv, \qv)$ as
an expansion in orders of $f$ so that
\begin{equation}
G_f (\pv, \qv) = \sum_{N=0}^{\infty} {G_f}^{(N)}(\pv, \qv)
\end{equation}
where $G_f^{(N)}(\pv, \qv)$ is of order $f^N$.

On the other hand, the standard 2 point function is 
\begin{equation}
G(\vec{X}, \vec{Y} )= {i\over (2\pi )^2} \int d^2k {e^{-i \kv\cdot (\vec{X} -
             \vec{Y})}\over \kprop}
\label{g2dxy}
\end{equation}
with Fourier transform
\begin{eqnarray}
G (\pv, \qv) & = & \int d^2x d^2y e^{-i\pv \cdot \xv } e^{i\qv \cdot \yv }
   G(\vec{X(\xv)}, \vec{Y(\yv)} ) ,\label{g2d0} \\ 
&=& i\delta (p_0,q_0) \int dx dy dk {e^{i(p+k)x} e^{-i(q+k)y}\over \kprop}
                                            e^{iq_0(f(x)-f(y))},
\label{g2d}
\end{eqnarray}
where we have substituted for $\vec{X}, \vec{Y}$ in terms of $\xv,\yv$. 
We can expand the last exponential in (\ref{g2d}) in a power series and hence
obtain $G$ as an expansion in powers of $f$ i.e.
\begin{equation}
G (\pv, \qv) = \sum_{N=0}^{\infty} {G}^{(N)}(\pv, \qv)
\label{g2dn}
\end{equation}
where $G^{(N)}(\pv, \qv)$ is of order $f^N$. 
 To second order we have, from (\ref{g2d}),
\begin{eqnarray}
G^{(0)}(\pv, \qv) &=& {i(2\pi)^2\over \pprop}\delta (\pv,\qv) \\
G^{(1)}(\pv, \qv)&=& 2\pi  q_0\delta (p_0,q_0) f(p-q) ({1\over \pprop}
                       -{1\over \qprop}) \\
G^{(2)}(\pv, \qv)&=& {-iq_0^2\over 2}\delta (p_0, q_0)
                     \int dk f(p+k) f(-q-k) \nonumber\\
& &({1\over \qprop} 
                             + {1\over \pprop} -{2\over \kqprop}) .
\end{eqnarray}
Clearly, $G^{(0)}(\pv, \qv) =G_f^{(0)}(\pv, \qv)$.
From (\ref{gf2d}) and (\ref{c1}), we have
\begin{eqnarray}
G^{(1)}_f(\pv, \qv) &= &{-i\over \pprop} C^{(1)} (\pv, \qv) {1\over \qprop} \nonumber\\
&=& {-i\over \pprop} 2\pi i \delta (p_0, q_0) f (p-q) p_0 (p^2-q^2)
         {1\over \qprop} \nonumber \\
&= & 2\pi p_0\delta (p_0, q_0)f (p-q) ({1\over \pprop} - {1\over \qprop})\nonumber\\
&=& G^{(1)} (\pv, \qv). 
\end{eqnarray}
Finally, from  (\ref{gf2d}), (\ref{c1}) and(\ref{c2})   we have
\begin{eqnarray} 
G_f^{(2)}(\pv, \qv)   &=&
         {i\over (2\pi)^2}{1\over \pprop}\int d^2k C^{(1)}(\pv, \kv)
          {1\over \kprop}C^{(1)}(\kv, \qv){1\over \qprop} \nonumber\\
  & +& {(-i)\over \pprop} C^{(2)}(\pv, \qv) {1\over \qprop} \nonumber \\
&=& -iq_0^2\delta (q_0, p_0)
\int dk \{{f(p-k) f (-q+k) \over (\pprop ) (\qprop )}\nonumber\\
& & \big(
          {(p^2- k^2) (k^2 -q^2) \over \kqprop} + (p-k)(k-q)\big)\}
\label{gf22d}
\end{eqnarray}
After some algebra it be shown that 
\begin{eqnarray}
G_f^{(2)}(\pv, \qv) =   G^{(2)}(\pv, \qv) & \nonumber\\
 +{iq_0^2 (p+q)\over (\pprop) (\qprop) }&\delta (q_0, p_0)
      \int dk f(p+k) f (-q-k) ({p+q\over 2}  +k) .&
\end{eqnarray}
 Setting ${p+q\over 2}  +k = l$ we have 
\begin{equation}
\int_{-\infty}^{\infty} dk f(p+k) f (-q-k) ({p+q\over 2}  +k)
=  \int_{-\infty}^{\infty}dl lf (l+{p-q\over 2}) f (-l +{p-q\over 2}),
\end{equation}
which vanishes by virtue of the integrand  being odd in $l$. Hence $G$ and 
$G_f$ are identical to 2nd order. We have exhibited the calculations
in some detail to show that this agreement is not entirely trivial
as well as to illustrate our ideas in a concrete setting. 
In the next section we discuss the case of an arbitrary foliation
in $n+1$ dimensions.

\subsection*{4.2 The general case.}

Integration of (\ref{lpi}) over $X^A$ and dropping of the irrelevant
 c- number determinant term  gives
\begin{equation}
Z = \int {\cal D} \phi \exp iS[\phi (x^i,t)] .
\end{equation}
Here  
\begin{equation}
S[\phi (x,t)] =-{1\over 2}
 \int d^{n+1}x{\sqrt{\eta}}(\eta^{\alpha \beta}\partial_{\alpha}\phi \partial_{\beta}\phi + m^2 \phi^2),
\label{actionf}
\end{equation}
where $\eta^{\mu \nu}$ is defined by (\ref{eta})  with $X^A=f^A$. 
In the above equation, the coordinates
$x^{\alpha}$ are {\em fixed once and for all} 
by the choice of the embedding $f^A(x,t)$. 
Just as for the 2 dimensional example in section 4.1, 
the action (\ref{actionf}) can be written as the sum of a free part and an 
interaction term describing interaction with external fields.

The free part of the action describes a scalar field propagating on a 
flat spacetime with {\em $x^{\alpha}$ as inertial coordinates} so that the 
line element of this spacetime is $ds^2= -(dt)^2 + \sum_{i=1}^n (dx^i)^2$. We denote the 
flat spacetime metric defined by this line element by 
$\eta^{\mu \nu}_f$. The analog  of 
$f(x,t)$ in section 4.1 is 
\begin{equation}
h^A (x^i, t) := X^A(x^i,t) - x^{\alpha}\delta^A_{\alpha}
\label{defha}
\end{equation}
i.e. $h^0 = T-t, h^1= X^1 - x^1$ etc. We define the external fields
 $f^{\mu \nu}$ and $\alpha$ by
\begin{equation}
f^{\mu \nu} = \sqrt{\eta} \eta^{\mu \nu} -\eta_f^{\mu \nu}, 
\;\;\;\;\; \alpha = \sqrt{\eta} -1 .
\end{equation}
The above equation is defined in the {\em fixed} $(x^i,t)$
coordinate system and  $\sqrt{\eta}$ is calculated in this coordinate system.
Note that in analogy to (\ref{fdef}),
 $f^{\mu \nu}$ and $\alpha$ can be obtained as a series expansion in powers of
$\partial_\alpha h^A$.

The action (\ref{actionf}) takes the form
\begin{equation}
S[\phi (x,t)] =-{1\over 2}
 \int d^{n+1}x(\eta_f^{\alpha \beta}\partial_{\alpha}\phi \partial_{\beta}\phi
 + m^2 \phi^2         
+ f^{\alpha \beta}\partial_{\alpha}\phi \partial_{\beta}\phi + \alpha m^2\phi^2) .
\label{pertf}
\end{equation}
$G_f$ can be defined as a standard perturbative quantum field theory 
expansion in powers of the external field 
$f^{\mu \nu}, \alpha$. 
This expansion makes use of the propagator defined from the
free part of the action which in turn derives its structure from the 
flat metric $\eta_f^{\mu \nu}$. 

Our general strategy is as follows.
The momentum space two point function, $G_f(\pv , \qv )$ is defined by 
an $n+1$ dimensional analog of the  expression (\ref{gf2d})
with 
$C( \kv , \lv)$ defined by an appropriate generalization of (\ref{cdef}). 
$C( \kv , \lv)$ itself is 
a sum over $C^{(N)}( \kv ,  \lv )$ where the $C^{(N)}$ are of order $(h^A)^N$.
Unlike the specific 2 
dimensional example discussed in section 4.1 where $N=1,2$, here $N$
can in general range from 1 to $\infty$. 
The $N$th order (in $h^A$)  
contribution $G_f^{(N)}$ to $G_f$ can be calculated. 
The standard  2 point function, $G(\Xv (\xv ), \Yv (\yv ))$ can
be Fourier transformed in analogy to (\ref{g2d0}) to give $G( \pv , \qv )$. 
The latter can be
expanded in powers of $h^A$ in analogy to the expansion defined by equation
(\ref{g2dn}). Finally, the $N$th order contributions $G^{(N)}(\pv , \qv )$
can be compared with $G_f^{(N)}(\pv ,  \qv )$.

Though the general strategy seems straightforward,
the following discussion indicates that there are complications in defining 
$G_f( \xv_1, \xv_2 )$ in the manner sketched above (similar complications,
arising from `illegal' expansions of the relevant exponential inside the
$n+1$ dimensional analog of (\ref{g2d}) for a general foliation, 
may exist for the computation of $G^{(N)} (\pv , \qv )$).
If $h^A (x^i, t)$ are of compact support in $x^{\alpha}$, one can check that 
contributions to $G_f (\pv , \qv )$ to  any order in 
$f^{\mu \nu}$ are UV finite.
For an arbitrary choice of $f^A$, $h^A$ is restricted by the boundary conditions (\ref{bdryx})  to be of compact support only in $x^i$ and not in both $x^i$ 
and $t$.  For generic choices of $h^A$, UV divergences may possibly
exist. Further, even if there are no UV divergences, $G_f(\xv_1, \xv_2 )$
is defined in position space via the inverse Fourier transform of 
$G_f ( \pv , \qv )$, the latter being the sum of contributions at every 
order of perturbation theory. Whether this sum converges well enough for its
inverse Fourier transform to exist (as a distribution) is also not clear.
We shall concern ourselves with an investigation
 of these issues in future work. 

Our computations in section 4.1 indicate that a brute force 
term by term analysis would probably be quite involved. In the remainder of 
this section we propose a line of attack springing from considerations 
of a more general nature.
In what follows,  we shall simply assume that there is some way to 
define $G_f ( \xv_1 , \xv_2 )$ as a distribution by using standard perturbative quantum field
theory techniques on the flat 
$\eta_f^{\mu \nu}$- spacetime such that its only singularities are when
$\xv_1 = \xv_2$. 
Once this fairly restrictive assumption has been made, 
$G_f ( \xv_1 , \xv_2 )$ is  constrained by the following argument.

The equations of motion from the action (\ref{actionf}) are
\begin{equation}
\partial_{\mu} (\sqrt{\eta}\eta^{\mu \nu}\partial_{\nu} \phi)
- \sqrt{\eta}m^2 \phi
= \sqrt{\eta} (\Box - m^2) \phi = 0 .
\end{equation}
Here $\Box - m^2 = \eta^{AB}\partial_A \partial_B - m^2$ is a {\em scalar}
differential operator. From its definition, it follows that $G_f( \xv_1, 
\xv_2 )$ is a Green's  function for the operator  
$\partial_{\mu} (\sqrt{\eta}\eta^{\mu \nu}\partial_{\nu})- \sqrt{\eta}m^2$ 
so that 
\begin{equation}
({\partial \over \partial x^{\mu}}
 \sqrt{\eta (x)}\eta^{\mu \nu}(x)
{\partial \over \partial x^{\nu}} - \sqrt{\eta}m^2) G_f (\xv , \yv ) = i
\delta (\xv , \yv).
\end{equation}
Using the fact that  $\delta (\xv , \yv)$ transforms as a unit density in its
first argument and as a scalar in its second, we have
\begin{equation}
(\Box - m^2) G_f (\xv , \yv ) = i
\delta (\Xv , \Yv).
\end{equation} 
Thus $G_f$ and $G$ are constrained by virtue of their being Green's functions for the same differential operator! In the context of our restrictive assumptions
this implies that their difference, $\Delta G_f (\xv_1, \xv_2)=
G(\Xv_1( \xv_1), \Xv_2(\xv_2 ))- G_f( \xv_1, \xv_2 ) $,  must be a smooth solution to the 
Klein Gordon equation.

First consider the case when $h^A$ are of compact support. This means that 
$x^{\alpha}$ agree with $X^A$ outside a compact region $K \subset R^{n+1}$
and that $f^{\mu \nu},\alpha$ vanish outside $K$.
Suppose that we could show that $\Delta G_f (\xv_1, \xv_2) =0$ for 
$\xv_1, \xv_2 \in R^{n+1} - K$(here 
$R^{n+1}-K$ refers to the complement of $K$). Then the following argument shows that 
$\Delta G_f (\xv_1, \xv_2) =0$ everywhere. 
Fix the point $p_1$ such that $p_1 \in R^{n+1}-K$.
Then, 
we have that
$\Delta G_f (\xv_1, \xv_2) = 0$ for $\xv_2 \in R^{n+1}-K$ and that 
$\Delta G_f (\xv_1, \xv_2)$ satisfies the Klein Gordon equation. From the 
uniqueness of evolution from initial data on a Cauchy slice contained
in $R^{n+1}-K$, it follows that $\Delta G_f (\xv_1, \xv_2)$ vanishes for
all $\xv_2 \in R^{n+1}$. Since $G$ and $G_f$ are symmetric in their
arguments, it follows that $\Delta G_f (\xv_1, \xv_2)$ vanishes for also for
all $\xv_1 \in R^{n+1}$ and all $\xv_2 \in R^{n+1} - K$. Again using uniqueness
of evolution from a Cauchy slice in $R^{n+1}-K$, it follows that 
$\Delta G_f (\xv_1, \xv_2)$ vanishes for all $\xv_1, \xv_2 \in R^{n+1}$.

If $h^A$ is not of compact support, it must be still be true from the 
boundary conditions (\ref{bdryx}) that $x^{\alpha}$ agrees with $X^A$ outside
a timelike tube $\tau$ and that $f^{\mu \nu}, \alpha$ vanish outside $\tau$.
Again, suppose that we could show that $\Delta G_f (\xv_1, \xv_2) =0$ for 
$\xv_1, \xv_2 \in R^{n+1} - \tau$.
Then, using the fact that 
the only smooth solution of the Klein- Gordon equation  with support restricted to $\tau$ is the trivial 
solution, \footnote{We shall
display our proof of this assertion 
elsewhere.}
 arguments similar to those used for 
the case of $h^A$  having compact spacetime support show that, once again,
$\Delta G_f (\xv_1, \xv_2)$ vanishes for all $\xv_1, \xv_2 \in R^{n+1}$.

Thus, the absence (or existence) of inequivalent quantizations has been 
reduced to the vanishing (or not) of  $\Delta G_f (\xv_1, \xv_2)$ for
$\xv_1, \xv_2$ both outside the support of $h^A$. We propose to analyse 
this behaviour of $\Delta G_f (\xv_1, \xv_2)$ through a {\em position space}
perturbative expansion of $G_f (\xv_1, \xv_2)$ in future work.

\section*{5. Discussion.}

The three issues dealt with in this work are the correct measure for the 
Lorentzian path integral, the construction of a convergent Euclidean path 
integral to compute the vacuum wave function and the possibility of 
inequivalent quantizations based on different choices of time.
These issues arise in the quantization of any generally covariant theory. 
Below, we remark on each of them in view of the results obtained in sections 
2 to 4.

The Lorentzian path integral measure obtained in appendix A1 is very similar to the one obtained by Fradkin and Vilkovisky in \cite{fv} (and anticipated
even earlier by Leutwyler in \cite{leut}) for quantum gravity. In both cases
(i.e. parametrised field theory and gravity), the measure appears non-covariant
in that it explicitly refers to the coordinate time $t$. In the case of 
gravity, Fradkin and Vilkovisky argue that their measure is, despite appearances to the contrary, diffeomorphism invariant. 
The key point is that diffeomorphisms corresponding to time reparametrizations
must be handled with extreme care because of the non-trivial contribution of 
non- smooth paths to the transition amplitude. The reason for the non-covariant
factors in the measure is that the path integral is defined in terms of 
the limit of a discretization which itself depends on the choice of time.
The non- covariant factor in the measure exactly compensates for this intrinsic
discretization dependent non- covariance, so as to make the measure
diffeomorphism invariant. Since parameterised field theory is also a 
space-{\em time} diffeomorphism invariant theory, the arguments in \cite{fv}
apply to it. Since we have derived the measure for this simple system from the 
correct Liouville measure in the phase space path integral, we believe 
\footnote{
We again emphasize that we have 
neglected endpoint terms and used only canonical gauges
in our arguments (see \cite{teitel,simeone}
in this regard). Neverthless, our intuition is that our result for the 
measure is sufficiently robust to survive  a more careful treatment  
of endpoint contributions.}
in its 
validity and interpret our results as supportive of the Fradkin- Vilkovisky
measure being the correct one  for quantum gravity as opposed to the 
more commonly used de Witt measure \cite{dewitt}.\footnote{
For perturbative quantum gravity calculations in a {\em dimensional
regularization scheme} contributions of the nontrivial local factor in 
the measure are regulated away to zero because they are 
proportional to 
$\delta^4(0)$. However for any non- perturbative treatment, this factor should
be important.} We also would like to note that contrary to what Fradkin and 
Vilkovisky state in \cite{fv2}, the above subtelities regarding the measure do
not have anything to do with the appearance of structure functions in the
constraint algebra; clearly in parameterised field theory the constraint
algebra for the constraints $C_A$ is abelian and no structure functions
appear. Rather, these subtelities seem to be entirely due to the 
property of general covariance.

Our definition of Euclideanization was motivated by the work of Schleich 
\cite{kristin}. She was interested in constructing diffeomorphism invariant, 
convergent Euclidean path integrals from the correct reduced phase space path
integral expression for the vacuum wave function. Her strategy was to start 
from the explicit reduced phase space path integral expression and rewrite it 
as a convergent, diffeomorphism invariant, configuration space path 
integral. For a treatment of gravity beyond perturbation theory, an explicit
characterization of the reduced phase space is not available and Schleich's
strategy is hard to implement. In this work we have suggested a strategy 
which does not require an explicit parameterisation of the reduced phase 
space. We start from a  gauge fixed expression in phase space and integrate
out the momenta. We are {\em not} concerned with maintaining diffeomorphism
invariance and indeed this is an aspect  of our constructions which we 
need to understand better. We have verified that the Euclidean action 
(\ref{le}) is spatially diffeomorphism invariant. We do not know if it
displays any sort of invariance related to Lorentzian time reparameterizations.
Although in general the Euclidean action is complex, it is easily verified 
that for the case of an 
inertial foliation, the action turns out to be the standard, real Euclidean 
action. The difference between Schleich's aims 
(as we understand them- of course we could be in error in our understanding) 
and 
ours may be stated in this context as follows. Whereas we are content with
the form of the action given by (\ref{le}), Schleich would take the flat 
foliation related standard Euclidean action and construct parameterised 
Euclidean field theory to obtain an explicitly diffeomorphism invariant,
convergent path integral expression for the vacuum wave function.

We would like to emphasize again that in our arguments for the phase
space path integral (equation (\ref{pqampa}) in section 3.1 
 and equation (\ref{redampa}) in appendix A2), we have neglected endpoint
contributions. These contributions are important (see \cite{ht} and 
\cite{teitel,simeone}). A more careful treatment of these endpoint 
contributions is desirable. Indeed, this seems to be the only possible obstacle
to an application of our ideas to quantum gravity. If we neglect the endpoint
contributions and  use a  canonical gauge independent of momenta, it does seem
possible to try out our proposal for asymptotically flat quantum gravity.
If endpoint contributions could be taken care of and if we could do the 
relevant calculations we would expect to get an, in general complex, convergent
Euclidean action for gravity. It is conceivable that progress could then be
made towards numerical evaluation of the vacuum wave function. In fact,
Loll and collaborators \cite{renate} have embarked on a program of 
numerical evaluation of
Euclidean path integrals but their analytical justification \cite{arundhati}
seems to have as an input, the de Witt measure, which we suspect is 
incorrect.

Finally we turn to a discussion of the issue of inequivalent quantizations. As
shown in section 4, we have connected this issue to that of gauge independence
of the time ordered 2 point function. Note that by virtue of our 
boundary conditions (\ref{bdryx})  we have disallowed all global Poincare 
transformations (with the exception of time translation). The 2 point function
in different gauges actually correspond to (in the Hamiltonian framework)
the evaluation of the vacuum expectation value of the same Dirac observable 
in different gauges. The Dirac observables can be constructed as 
``evolving constants of motion'' \cite{evolvconst} from observables 
corresponding to initial data on a fixed $T=0$ slice. The latter observables
can be constructed from the data $(\phi, \pi, X^A, P^A)$ by a Hamilton- Jacobi
type of canonical transformation \cite{kkiyer}.
Note that the sort of gauge independence which would ensure the absence of 
inequivalent quantizations is qualitatively different from the more 
commonly encountered gauge independence of the S- matrix in Poincare 
invariant quantum field theory. It may well turn out that there is no
inequivalent quantization as far as the 2 point function is concerned but,
as we have tried to argue, the verification of this is non-trivial.
We would also like to make contact with the existence of unitarily 
inequivalent quantizations in higher dimensions noted in \cite{ctmv2} in the context of 
canonical quantization.

\noindent{\bf Acknowledgements:}
We are indebted to Ghanashyam Date for numerous discussions. 
We thank Sebastian Guttenberg and Joseph Samuel
for a careful reading of a preliminary version 
 of this work and their comments.
We would also like to
thank Abhishek Dhar, Harendranath,  Joseph Samuel and Sumati Surya 
for very useful discussions.

\section*{Appendix}
\subsection*{A1}
Consider the following
 momenta-independent but otherwise arbitrary gauge fixing conditions
$\chi_A [X^B, \phi;y)= 0$.
The notation indicates that $\chi^A$ is a functional of $X^B$ and $\phi$
and a function of the point with coordinates $y^{\alpha}$. In what follows 
we shall suppress the $X^B,\phi$ dependence in our notation. 
The contribution of the ghosts $(\omega^A, \omega^{*A})$ to the phase space
action (\ref{hamaction}) is
\begin{equation}
S_{gh}= \int d^{n+1}x d^{n+1}y \omega^{*A}(x)\{C_A(x), \chi_B (y) \}\omega^B(y)
\end{equation}
where
\begin{equation}
\{C_A(x), \chi_B (y) \}= -{\delta \chi_B(y)\over \delta X^A (x)}
    - {\delta \chi_B(y)\over \delta \phi (x)}(
-n_A(x) {\pi (x)\over \sqrt{q(x)}} + q^{ij}(x)X_{Aj}(x)\partial_j \phi (x)) .
\label{cagf}
\end{equation}
Integration of the phase space path integral over $M^A, P_A$ can be done as 
before to obtain
\begin{equation}
Z = \int {\cal D} \phi{\cal D}\pi {\cal D}X{\cal D}\omega^{*}
{\cal D}\omega
      \delta [\chi ]
    \exp (i\int dtd^nx (\pi {\dot{\phi}} -Nh -N^ih_i) +iS_{gh})
\label{hampia2}
\end{equation}
Notice that from (\ref{cagf}), $S_{gh}$ is linear in $\pi$ and hence
the total action is 
still quadratic in $\pi$. It is straightforward to integrate
(\ref{hampia2}) over $\pi$ to obtain
\begin{equation}
Z = \int {\cal D} \phi{\cal D}X{\cal D}\omega^{*}
{\cal D}\omega [{\rm det}{iN\over \sqrt{q}}]^{-{1\over 2}}
      \delta [\chi ]
    \exp (i\int dtd^nx (\pi {\dot{\phi}} -Nh -N^ih_i) +iS_{gh})
\label{hampila}
\end{equation}
where $S_{gh}$ is evaluated at the classical value of $\pi$ given by  
\begin{equation}
\pi_{class} = \sqrt{q} 
({\dot{\phi} -N^i\partial_i\phi \over N}).
\label{pidef}
\end{equation}
The only non-trivial step in the
computation is to use the Grassmanian nature of $\omega^{*A}$ to conclude that
$(\omega^{*A}(x) n_A(x))^2$ vanishes. As usual, $q^{ij} , N, N^i$ are
interpreted as functions of $X^A$. The ghost variables can be integrated
over to give the determinant of $\{ C_A(x), \chi_B(y)\}$ where the latter
is given by the right hand side of equation 
 (\ref{cagf}) evaluated at $\pi= \pi_{class}$ given by 
(\ref{pidef}). It is straightforward to verify that 
\begin{equation}
\{C_A(x), \chi_B (y) \}|_{\pi =\pi_{class}}
= -\big( 
{\delta \chi_B(y)\over \delta X^A (x)}
    +{\delta \chi_B(y)\over \delta \phi (x)} 
{\partial \phi (x)\over \partial X^A(x)}\big) .
\label{cagfl}
\end{equation}
The operator ${\partial \;\;\;\over \partial X^A(x)}$ is defined via the
invertible dependence of the embeddings $X^A(x)$ on the coordinates 
$x^{\alpha}$. Thus, equation (\ref{cagfl}) can be rewritten as 
\begin{eqnarray}
\{C_A(x), \chi_B (y) \}|_{\pi =\pi_{class}} & = &
-{\delta {\cal L}_{\xi}\chi_B (y)\over \delta \xi^A (x)} \nonumber \\
&= & -{\delta {\cal L}_{\xi}\chi_B (y)\over \delta \xi^{\alpha} (x)}
      {\partial x^{\alpha}\over \partial X^A (x)} . \\
\label{fpa}
\end{eqnarray}
Using this and the fact that the Jacobian of the coordinate transformation
from $X^A \rightarrow x^{\alpha}$ is $\sqrt {\eta}$, we 
have
\begin{equation}
Z = \int d\mu[\phi, X^A]
      \delta [\chi_B] {\rm det}\big({\delta {\cal L}_{\xi}\chi_B (y)\over \delta \xi^{\alpha} (x)}\big)
    \exp iS[\phi, X^A] .
\label{pila}
\end{equation}
Here $S[\phi, X^A]$ is the classical action (\ref{actionpft}), and the
path integral measure can be written in the context of an appropriate 
discretization  as 
\begin{equation}
d\mu[\phi, X^A] =
\prod_{x} \eta^{tt}(x)\eta^{-{1\over 4}}  (x)
d\phi (x)dX^A (x).
\label{measure}
\end{equation}

\subsection*{A2.}

In the case where the gauge fixing constraints are independent of momenta,
gauge independence of the transition amplitude
(\ref{pqamp})
may  be shown (modulo the caveats mentioned in section 3.1)
through  methods similar to those
employed in \cite{sundermeyer}.
We make a canonical transformation from $(q_i, p_i), i=1...n$, to
new conjugate pairs
$({\bar q}_l, {\bar p}_l), l=1...n-m$ and  
$(Q_{\alpha}, P_{\alpha}), \alpha=1...m$.
Here $Q_{\alpha}= \chi_{\alpha}$ and $({\bar q}_l, Q_{\alpha})$  
encode the same information as $q_i$.
On the surface defined by $Q_{\alpha}= C_{\alpha}=0$,
$P_{\alpha}$ is a function of 
${\bar q}_l, {\bar p}_l$. It can be checked  that  
(\ref{pqamp}) reduces to 
\begin{equation}
Z({\bar q}_{lI}, t_I;{\bar q}_{lF}, t_F) 
= \int {\cal D}{\bar q}{\cal D}{\bar p}
\exp (i\int {\bar p}_l {\dot {\bar q}}_l - H({\bar p},{\bar q})).
\label{redamp}
\end{equation}
Gauge independence of (\ref{pqamp}) under infinitesmal changes of gauge can be 
checked \cite{sundermeyer}
by subjecting the gauge condition to a canonical transformation
generated by the constraints. In this treatment, 
endpoint contributions
arising from the canonical transformations encountered are ignored
and (\ref{pqamp}) (as well as (\ref{redamp})) 
is identified with the transition amplitude between the
gauge fixed endpoints 
$q_{iI}$ and $q_{iF}$. 

An identical treatment can also be applied to 
show the gauge independence of
the expression (\ref{pqampa}).
It is straightforward to check that (\ref{pqampa}) reduces to
\begin{equation}
Z_{a}({\bar q}_{lI}, t_I;{\bar q}_{lF}, t_F ) = 
\int {\cal D}{\bar q}{\cal D}{\bar p}
\exp (i\int {\bar p}_l {\dot {\bar q}}_l - aH({\bar p},{\bar q})),
\label{redampa}
\end{equation}
and that its
gauge independence is ensured by virtue of the fact that 
$H(q_i, p_i)$ commutes (weakly) with the constraints. Again, we disregard
various endpoint contributions coming from canonical transformations.
It is straightforward to
see that in operator language, 
\begin{equation}
Z_a = <{\bar q}_{lF}, t_F | \exp (-i(a-1){\hat H}(t_F-t_I))| 
{\bar q}_{lI}, t_I>.
\end{equation}
Since $(q_{lI}, q_{lF})$ satisfy the gauge conditions, we may 
identify them with $({\bar q}_{lI}, {\bar q}_{lF})$. 
Then,
 under the assumption that the ground state energy vanishes and with $a$ chosen
such that it has negative imaginary part,
the 
usual Feynman-Kac type arguments  show that 
\begin{equation}
Z_{a}({\bar q}_{lI}, t_I=-\infty ;{\bar q}_{lF}, t_F )
=Z_{a}(q_{iI}, t_I=-\infty ;q_{iF}, t_F)
= \psi_0 (q_F,t_F) \psi_0^*(q_I,t_I) .
\label{psi0app}
\end{equation}
Here $\psi_0$ is the vacuum wave function, the vacuum being defined as the 
lowest energy state of the quantum operator corresponding to the classical
Hamiltonian $H$.


\end{document}